# Revisiting Theoretical Modeling of Seebeck Coefficient of Semiconductors

Yu-Chao Hua*

Here, we revisit the closed-form models of Seebeck coefficient, and identify the thermodynamic flaws in those extensively-used methodologies, essentially including the confusion of electrochemical potential and electrical potential within the definition, arbitrarily neglecting the dependence on conduction band bottom and Fermi level when calculating distribution function's spatial gradient, and the improper heat flux presentation. Regarding these issues, an alternative methodology is presented, which derives the Seebeck coefficient model based on the drift-diffusion equation with the Soret effect in the open circuit condition. The reversible model that eliminates the electron velocity and relaxation terms in the formulation is recovered for the band term in the near-equilibrium case, and it can give the lower bound of Seebeck coefficient in theory, while the phonon-drag term is of the identical form to the Boltzmann transport equation (BTE)-based one. A case study is performed for highly-n-doped Silicon, where the reversible, ballistic (based on Landauer formulation) and BTE models are compared with each other and with the experimental data. The reversible model gives the fairly-good predictions for the experiments. Given its theoretical-soundness, simple form and low computation cost, the reversible model can serve as a concise alternative for evaluating thermoelectric properties in both the reversible and near-equilibrium cases.

*Corresponding author: Email: huayuchao19@163.com

# I. Introduction

Research on thermoelectricity (TE) of semiconductors has long been an active area due to the perspective of developing doped-semiconductors-based solid-state electricity generators [1][2] and its theoretical importance as a paradigmatic model in irreversible thermodynamics [3][4][5][6][7][8][9]. Particularly, Seebeck coefficient is one of the central parameters in the TE studies, which characterizes the open circuit (OC) voltage (i.e., Seebeck voltage) induced by a temperature difference.

The Seebeck coefficient ($\alpha_{\mathrm{S}}$) is defined as [10],

$$\alpha_{\mathrm{S}} = -\frac{V_{\mathrm{h}} - V_{\mathrm{c}}}{T_{\mathrm{h}} - T_{\mathrm{c}}} = \frac{1}{q_{\mathrm{e}}} \frac{E_{\mathrm{F_h}} - E_{\mathrm{F_c}}}{T_{\mathrm{h}} - T_{\mathrm{c}}} = \frac{1}{q_{\mathrm{e}}} \frac{\Delta E_{\mathrm{F}}}{\Delta T}, \tag{1}$$

where the subscripts “h” and “c” refers to the hot and cold terminals respectively, $V_{\mathrm{h(c)}}$ is the measured value by voltmeter, and $T_{\mathrm{h(c)}}$ is temperature. It is important to emphasize what measured by a voltmeter is the difference of electrochemical potential (i.e., Fermi level $E_{\mathrm{F}}$ in semiconductors) divided by charge quantity ($q_{\mathrm{e}}$, negative for electrons and positive for holes) rather than the electrostatic potential difference [10][11]. For instance, Figure 1 shows the band diagrams of electron system (n-doped semiconductors) at (a) low temperature $T_c$ and (b) high temperature $T_h$, respectively. The Fermi level is given by,

$$E_F = \mu + E_C = \mu + q_e\varphi, \tag{2}$$

in which $\mu$ is chemical potential, $E_C$ the conduction band bottom, and $\varphi$ is electrostatic potential. Note that the Fermi level will shift to the middle of band gap with temperature increasing, as illustrated in Figs. 1(a) and (b). Note that the Seebeck coefficient generally contains the band term (dominated in highly/moderately-doped and high temperature cases) and the drag term (like phonon drag that becomes important in low-doped and low temperature regions) [12][13]. The band term is usually more important for the practical TE generators, since they generally require highly/moderately-doped semiconductors [14][15].

The theoretical modeling of Seebeck coefficient, especially to derive its closed-form predictive model, provides benchmark and guidelines for mechanisms analysis and TE material R&D [14]. Currently, Boltzmann transport equation (BTE) is the major theoretical tool to study the Seebeck coefficient [16][17][18][19]. Under the relaxation time approximation (RTA) and local equilibrium assumption (LEA), a closed-form BTE-based model of the band term can be derived [20],

$$\alpha_{S(BTE)} = \frac{1}{q_e T}\frac{\int \tau_e v_e^2 f_{FD}[1-f_{FD}][\varepsilon_{\mathbf{k}} - E_F]D_{OS}(\varepsilon_{\mathbf{k}})d\varepsilon_{\mathbf{k}}}{\int \tau_e v_e^2 f_{FD}[1-f_{FD}]D_{OS}(\varepsilon_{\mathbf{k}})d\varepsilon_{\mathbf{k}}}, \tag{3}$$

with Fermi-Dirac statistics,

$$f_{FD} = \frac{1}{e^{\left(\frac{\varepsilon_{\mathbf{k}}-\mu}{k_B T}\right)}+1} = \frac{1}{e^{\left(\frac{\varepsilon_{\mathbf{k}}-[E_F-E_C]}{k_B T}\right)}+1}, \tag{4}$$

in which $\tau_e$ is the relaxation time of electrons, $v_e$ is the electron velocity, $\varepsilon_{\mathbf{k}}$ is energy and $Dos(\varepsilon_{\mathbf{k}})$ is the density of states (DOS). Besides, the phonon-drag term model is given by,

$$\alpha_{\mathrm{S(d)}} = \frac{\int v_{\mathrm{e}}\tau_{\mathrm{e}}D_{\mathrm{OS}}(\varepsilon_{\mathbf{k}})d\varepsilon_{\mathbf{k}}\,\beta_{\mathrm{ph}\to\mathrm{e}}}{\int \tau_e v_e^2 f_{\mathrm{FD}}[1-f_{\mathrm{FD}}]D_{\mathrm{OS}}(\varepsilon_{\mathbf{k}})d\varepsilon_{\mathbf{k}}}, \tag{5}$$

where $\beta_{\mathrm{ph}\to\mathrm{e}}$ the probability density of non-equilibrium population per energy interval excited by non-equilibrium phonons that can be calculated through the Fermi's golden rule [13].

There are generally two conventional schemes to obtain the models above: one is based on the OC condition where the net charge flux vanishes due to the balance between the temperature-gradient-driven and the external-force-driven parts; another one is to characterize the ratio between the heat flux and the charge flux (i.e. Peltier coefficient) in the isothermal condition case along with the Kelvin relation [21][22].

In fact, both of them have flaws in the view of thermodynamics. As for the 1st scheme, some papers often confuse the electrochemical potential and electrostatic potential and thus calculate the Seebeck coefficient using the electrostatic potential in the BTE [16][17]; in some other papers [18], the electrochemical potential is directly introduced into the BTE for well defining the Seebeck coefficient, which may contradict the conventional BTE of which external force term should correspond to the electrical field alone [23]; additionally, when evaluating the spatial gradient of distribution function in the BTE, most of work [17][18] merely considers the temperature-variation term and almost arbitrarily neglects those of conduction band bottom and Fermi level that both change across the sample even in the OC condition as shown in Fig. 1.

Regarding the 2nd scheme, the major concern comes from the definition of local heat flux ($J_Q$) that is characterized as the product of temperature and local entropy flux ($J_S$) [7],

$$J_Q = TJ_S = J_E - \mu J_N, \tag{6}$$

With energy flux $J_E$ and particle flux $J_N$. This is a rather strong assumption that not only needs the LEA but also requires assuming the local element undergoes a quasi-static (reversible) process where no local entropy production exists [7], which leads to a paradox with the fundamental principle of non-equilibrium thermodynamics [24]. Although this heat flux representation has been extensively adopted in the TE researches [19], it has been rarely clarified carefully in the literatures, which should arise the question whether the model based on it correctly corresponds to the physical essence of Peltier coefficient (defined as the ratio between heat flux and charge flux). Besides the BTE-based methodology, such heat flux representation has also been applied in the Landauer-Büttiker theory (in the regime of far-nonequilibrium ballistic transport, and it gives the ballistic model of Seebeck coefficient) [25], the Kubo formula [26][27], and other theoretical schemes [28][29], when dealing with the Seebeck coefficient, which can all face the same concern.

Additionally, some efforts have been paid to derive the predictive models of Seebeck coefficient through alternative methodologies with excluding the explicit heat flux presentation: Lei used the carrier hydrodynamic balance equation [30] to characterize the band term in the OC condition; Apertet et al. employed the drift-diffusion equation [10] to deal with the band term for the non-degenerate semiconductors. Moreover, the equilibrium thermodynamics can be used to characterize the band term of Seebeck coefficient through designing the corresponding quasi-static process [5][31], which can give the reversible model that holds concise form and validity even for strong-coupling systems but eliminates influence of irreversibility (the group velocity and relaxation time, which closely relate to irreversible mechanisms, are dropped)[6].

After revisiting the theoretical modeling of Seebeck coefficient, it is identified that some flaws remain in the view of thermodynamics, even for the extensively-used methodologies. In the present work, we employ the drift-diffusion equation with the Soret effect based on the OC condition to derive the closed-form models for both the band term and the phonon-drag term of Seebeck coefficient, which avoids the ambiguity between the electrochemical potential and electrical potential and excludes the explicit heat flux presentation. It is found that in the near-equilibrium case where the LEA is valid, that reversible model, which has been presented in our previous paper [6], can also work for the band term of Seebeck coefficient, even though it drops the velocity and relaxation terms, while the phonon-drag term keeps the same form as the BTE-based phonon-drag model. Importantly, the reversible model can provide a meaningful lower bound of Seebeck coefficient in theory. Furthermore, a case study is performed for highly-n-doped Silicon, where the reversible, ballistic and BTE models are compared with the experimental data and with each other. The reversible model gives the fairly-good estimations compared to the experiments, while it is theoretically-sound and of simple form and low computation cost.

# II. Theoretical Modeling of Seebeck Coefficient

## *A. Derivation based on OC condition in near-equilibrium case*

This section investigates the band term of Seebeck coefficient, while the phonon-drag term is handled afterwards.

In experiments, the Seebeck coefficient is frequently measured in an OC [22]. Figure 2 (a) shows the band diagram configuration when two heat sinks of low and high temperatures are connected by a n-doped semiconductor bridge in the OC condition. In the near-equilibrium case, the LEA assumes even if a system is not in global equilibrium, the basic functionals developed in the thermodynamic equilibrium can be inherited in small regions of it and enables us evaluate the state variables, like temperature (*T*), locally. Here, we assume the uniform doping case and thus result in the local electrical neutrality,

$$n_{\mathrm{d}}^{+} - n = 0, \tag{7}$$

where $n_{\mathrm{d}}^{+}$ is the local density of ionized donners and *n* is the local electron density. Combining the electric neutrality with Poisson's equation yields the linear distribution of electrostatic potential $\varphi$ within the conductor. In terms of Fourier's law-based heat conduction equation without sources, the temperature distribution is linear. Then, the distribution of Fermi level within the conductor is given by,

$$E_{\mathrm{F}}(x) = \mu(x) + E_{\mathrm{C}}(x) = E_{\mathrm{F_c}} + \frac{\partial E_{\mathrm{F}}}{\partial T}(T(x) - T_{\mathrm{c}}). \tag{8}$$

We could ignore the temperature dependence of $\partial E_{\mathrm{F}}/\partial T$, when the temperature difference is not that big, and thus $E_{\mathrm{F}}(x)$ is also a linear function. Hence, the Seebeck coefficient can have an equivalent local form,

$$\alpha_{\mathrm{S}} = \frac{1}{q_{\mathrm{e}}}\frac{\partial E_{\mathrm{F}}}{\partial T}. \tag{9}$$

In this way, instead of deriving $\Delta E_{\mathrm{F}}$, we can model the Seebeck coefficient locally under the LEA.

To derive $\partial E_{\mathrm{F}}/\partial T$, we need use the condition of vanishing net charge flux in the OC condition, using the drift-diffusion equation with the Soret effect which is given by,

$$J_{\mathrm{DD}} = -D_{\mathrm{e}}\frac{\partial n}{\partial x} - n\mu_{\mathrm{e}}\frac{\partial E_{\mathrm{C}}}{\partial x} - nD_{\mathrm{T}}\frac{\partial T}{\partial x}, \tag{10}$$

with diffusivity $D_{\mathrm{e}}$, mobility $\mu_{\mathrm{e}}$, and thermal diffusivity $D_{\mathrm{T}}$. Using the detailed balance condition, we have the generalized Einstein relations,

$$n\mu_{\mathrm{e}} = -D_{\mathrm{e}}\frac{\partial n}{\partial E_{\mathrm{C}}} = D_{\mathrm{e}}\frac{\partial n}{\partial \mu}, \tag{11}$$

and

$$nD_{\mathrm{T}} = -D_{\mathrm{e}}\frac{\partial n}{\partial T}. \tag{12}$$

In terms of the zero net charge flux condition, Eq. (10) becomes,

$$J_{\mathrm{DD}} = -D_{\mathrm{e}}\frac{\partial n}{\partial x} - n\mu_{\mathrm{e}}\frac{\partial E_{\mathrm{C}}}{\partial x} - nD_{\mathrm{T}}\frac{\partial T}{\partial x} = 0. \tag{13}$$

Using the generalized Einstein relations above and the equation,

$$\frac{\partial E_{\mathrm{C}}}{\partial x}=-\frac{\partial \mu}{\partial x}+\frac{\partial E_F}{\partial T}\frac{\partial T}{\partial x}, \tag{14}$$

it yields,

$$0=-D_{\mathrm{e}}\frac{\partial n}{\partial x}-D_{\mathrm{e}}\frac{\partial n}{\partial \mu}\left(-\frac{\partial \mu}{\partial x}+\frac{\partial E_F}{\partial T}\frac{\partial T}{\partial x}\right)+D_{\mathrm{e}}\frac{\partial n}{\partial T}\frac{\partial T}{\partial x}=D_{\mathrm{e}}\frac{\partial T}{\partial x}\left(\frac{\partial n}{\partial \mu}\frac{\partial E_F}{\partial T}-\frac{\partial n}{\partial T}\right). \tag{15}$$

We thus obtain,

$$\frac{\partial E_F}{\partial T}=\frac{\partial n}{\partial T}\bigg/\left(\frac{\partial n}{\partial \mu}\right). \tag{16}$$

When approximating $n$ using the FD distribution, i.e. $n_{\mathrm{FD}}$, following the LEA, the reversible model [6] of Seebeck coefficient is recovered in terms of Eqs. (1) and (16), which is given by,

$$\alpha_{\mathrm{S(R)}}=\frac{1}{q_{\mathrm{e}}}\frac{\int \frac{\partial f_{\mathrm{FD}}}{\partial T}D_{\mathrm{OS}}(\varepsilon_{\mathbf{k}})d\varepsilon_{\mathbf{k}}}{\int \frac{\partial f_{\mathrm{FD}}}{\partial \mu}D_{\mathrm{OS}}(\varepsilon_{\mathbf{k}})d\varepsilon_{\mathbf{k}}}=\frac{1}{q_{\mathrm{e}}T}\frac{\int f_{\mathrm{FD}}[1-f_{\mathrm{FD}}][\varepsilon_{\mathbf{k}}-E_{\mathrm{F}}]D_{\mathrm{OS}}(\varepsilon_{\mathbf{k}})d\varepsilon_{\mathbf{k}}}{\int f_{\mathrm{FD}}[1-f_{\mathrm{FD}}]D_{\mathrm{OS}}(\varepsilon_{\mathbf{k}})d\varepsilon_{\mathbf{k}}}. \tag{17}$$

Peterson and Shastry [27] presented an alike expression (named Kelvin formula) as Eq. (17) by slow-limit simplification of the Kubo formula, and regarded it as an approximate expression for the exact band term of Seebeck coefficient. Actually, in terms of our clarification above, this reversible model, which holds very concise form, can be a fairly-good estimation for the band term of Seebeck coefficient in the case of near-equilibrium, though it neglects the influence of scatterings. This is reasonable indeed that the Seebeck coefficient becomes its reversible version under the LEA that assumes the local functionals following their equilibrium-thermodynamic counterparts.

## *B. Phonon-drag term*

Due to the contribution of phonon drag effect, the net charge flux becomes,

$$J_{\mathrm{DDd}}=-D_{\mathrm{e}}\frac{\partial n}{\partial x}-n\mu_{\mathrm{e}}\frac{\partial E_{\mathrm{c}}}{\partial x}-nD_{\mathrm{T}}\frac{\partial T}{\partial x}+J_{\mathrm{d}}, \tag{18}$$

where $J_{\mathrm{d}}$ is the charge flux owing to the momentum exchange from the non-equilibrium phonons. It can be formulated as [13],

$$J_{\mathrm{d}}=\int v_{\mathrm{e}}\tau_{\mathrm{e}}D_{\mathrm{OS}}(\varepsilon_{\mathbf{k}})d\varepsilon_{\mathbf{k}}\,\beta_{\mathrm{ph}\to\mathrm{e}}\frac{\partial T}{\partial x}=L_{\mathrm{ph}\to\mathrm{e}}\frac{\partial T}{\partial x}, \tag{19}$$

Substituting Eq. (19) into Eq. (18) yields,

$$J_{\mathrm{DDd}}=-D_{\mathrm{e}}\frac{\partial n}{\partial x}-n\mu_{\mathrm{e}}\frac{\partial E_{\mathrm{c}}}{\partial x}-nD_{\mathrm{T}}\frac{\partial T}{\partial x}+L_{\mathrm{ph}\to\mathrm{e}}\frac{\partial T}{\partial x}, \tag{20}$$

Following the same procedure in Sec. II. A, we have,

$$J_{\mathrm{DDd}}=D_{\mathrm{e}}\frac{\partial T}{\partial x}\left(\frac{\partial n}{\partial \mu}\frac{\partial E_F}{\partial T}-\frac{\partial n}{\partial T}\right)+L_{\mathrm{ph}\to\mathrm{e}}\frac{\partial T}{\partial x}=0, \tag{21}$$

and

$$0=\frac{\partial E_F}{\partial T}-\frac{\partial n}{\partial T}\Big/\left(\frac{\partial n}{\partial \mu}\right)+\frac{L_{\mathrm{ph}\to\mathrm{e}}}{D_{\mathrm{e}}\left(\frac{\partial n}{\partial \mu}\right)}=\frac{\partial E_F}{\partial T}-q_{\mathrm{e}}\alpha_{\mathrm{S(R)}}+\frac{L_{\mathrm{ph}\to\mathrm{e}}}{D_{\mathrm{e}}\left(\frac{\partial n}{\partial \mu}\right)}. \tag{22}$$

We split $\partial E_{\mathrm{F}}/\partial T$ into two parts,

$$\frac{\partial E_F}{\partial T}=\left(\frac{\partial E_F}{\partial T}\right)_{\mathrm{R}}+\left(\frac{\partial E_F}{\partial T}\right)_{\mathrm{d}}, \tag{23}$$

and take $\left(\frac{\partial E_F}{\partial T}\right)_{\mathrm{R}}=q_{\mathrm{e}}\alpha_{\mathrm{S(R)}}$. Then, the phonon-drag term is derived,

$$\alpha_{\mathrm{S(d)}}=\frac{1}{q_{\mathrm{e}}}\left(\frac{\partial E_F}{\partial T}\right)_{\mathrm{d}}=-\frac{L_{\mathrm{ph}\to\mathrm{e}}}{q_{\mathrm{e}}D_{\mathrm{e}}\left(\frac{\partial n}{\partial \mu}\right)}. \tag{24}$$

Furthermore, according to Eq. (11), we reach,

$$\alpha_{\mathrm{S(d)}} = \frac{L_{\mathrm{ph\to e}}}{q_{\mathrm{e}} n \mu_{\mathrm{e}}}, \tag{25}$$

which is of the identical form to the BTE-based phonon drag model, Eq. (5), when combining the BTE-based model of mobility[18].

*C. Derivation based on OC condition in ballistic case*

This section deals with the Seebeck coefficient in the ballistic case based on the OC condition. The essential difference in the ballistic transport compared to the near-equilibrium case lies in the invalidity of LEA within the conductor bridge. Since the temperature difference $\Delta T$ is given, the Seebeck coefficient is calculated through the Fermi level difference $\Delta E_{\mathrm{F}}$ between the terminals (as illustrated in Fig. 2(b)), in terms of Eq. (1). In this case, the zero net charge flux is realized by the balance between the charge flux from the hot and cold terminals.

The Landauer formulation is used to calculate the charge flux from the hot terminal,

$$J_{\mathrm{Ba_h}} = \int_{E_{\mathrm{C_h}}}^{\infty} v_{\mathrm{e}} f_{\mathrm{FD}}\left(T_{\mathrm{h}}, E_{\mathrm{C_h}}, E_{\mathrm{F_h}}\right) D_{\mathrm{OS}}(\varepsilon_{\mathbf{k}}) d\varepsilon_{\mathbf{k}}. \tag{26}$$

The integration lower bound is shifted upwards by $E_{\mathrm{C_h}}$,

$$J_{\mathrm{Ba_h}} = \int_{0}^{\infty} \frac{v_{\mathrm{e}} D_{\mathrm{OS}}(\varepsilon_{\mathbf{k}})}{e^{\left(\frac{\varepsilon_{\mathbf{k}} - E_{\mathrm{F_h}}}{k_{\mathrm{B}} T_{\mathrm{h}}}\right)} + 1} d\varepsilon_{\mathbf{k}}. \tag{27}$$

Similarly, the charge flux from the cold terminal is expressed as,

$$J_{\mathrm{Ba_c}} = \int_{E_{\mathrm{C_c}}}^{\infty} v_{\mathrm{e}} f_{\mathrm{FD}}\left(T_{\mathrm{c}}, E_{\mathrm{C_c}} E_{\mathrm{F_c}}\right) D_{\mathrm{OS}}(\varepsilon_{\mathbf{k}}) d\varepsilon_{\mathbf{k}} = \int_{0}^{\infty} \frac{v_{\mathrm{e}} D_{\mathrm{OS}}(\varepsilon_{\mathbf{k}})}{e^{\left(\frac{\varepsilon_{\mathbf{k}} - E_{\mathrm{F_c}}}{k_{\mathrm{B}} T_{\mathrm{c}}}\right)} + 1} d\varepsilon_{\mathbf{k}} \tag{28}$$

The net flux is thus given by,

$$J_{\mathrm{Ba}}=\int_0^{\infty}\frac{v_{\mathrm{e}}D_{\mathrm{OS}}(\varepsilon_{\mathbf{k}})}{e^{\left(\frac{\varepsilon_{\mathbf{k}}-E_{\mathrm{F_h}}}{k_{\mathrm{B}}T_{\mathrm{h}}}\right)}+1}d\varepsilon_{\mathbf{k}}-\int_0^{\infty}\frac{v_{\mathrm{e}}D_{\mathrm{OS}}(\varepsilon_{\mathbf{k}})}{e^{\left(\frac{\varepsilon_{\mathbf{k}}-E_{\mathrm{F_c}}}{k_{\mathrm{B}}T_{\mathrm{c}}}\right)}+1}d\varepsilon_{\mathbf{k}}. \tag{29}$$

As $T_{\mathrm{h}}-T_{\mathrm{c}}=\Delta T\rightarrow 0$, it yields,

$$J_{\mathrm{Ba}}=\int_0^{\infty}v_{\mathrm{e}}\left\{\frac{\partial f_{\mathrm{FD}}}{\partial T}\Delta T+\frac{\partial f_{\mathrm{FD}}}{\partial E_{\mathrm{F}}}\Delta E_{\mathrm{F}}\right\}D_{\mathrm{OS}}(\varepsilon_{\mathbf{k}})d\varepsilon_{\mathbf{k}}. \tag{30}$$

With the net charge flux equal to zero, Eq. (30) leads to the conventional ballistic Seebeck coefficient in the literatures [25],

$$\alpha_{\mathrm{S(Ba)}}=\frac{1}{q_{\mathrm{e}}T}\frac{\int v_{\mathrm{e}}f_{\mathrm{FD}}[1-f_{\mathrm{FD}}][\varepsilon_{\mathbf{k}}-E_{\mathrm{F}}]D_{\mathrm{OS}}(\varepsilon_{\mathbf{k}})d\varepsilon_{\mathbf{k}}}{\int v_{\mathrm{e}}f_{\mathrm{FD}}[1-f_{\mathrm{FD}}]D_{\mathrm{OS}}(\varepsilon_{\mathbf{k}})d\varepsilon_{\mathbf{k}}}. \tag{31}$$

*D. Theoretical bounds of band term for 3D materials*

Since the phonon-drag term derived in our proposed theoretical scheme maintains the same form as the conventional one, we here focus on comparison of the varied formulations of band term, i.e., Eq. (3) (BTE model), Eq. (17) (Reversible model), Eq. (31) (Ballistic model). Within these three models, the DOS $D_{\mathrm{OS}}$, the electron velocity $v_{\mathrm{e}}$, and the relaxation time $\tau_{\mathrm{e}}$ can all be expressed in the form of power function dependent on the energy $\varepsilon_{\mathbf{k}}$ [23]. Hence, they can be expressed as a unified form,

$$\alpha_{\mathrm{S}}=\frac{1}{q_{\mathrm{e}}T}\frac{\int \varepsilon_{\mathbf{k}}^{r}f_{\mathrm{FD}}[1-f_{\mathrm{FD}}][\varepsilon_{\mathbf{k}}-E_{\mathrm{F}}]d\varepsilon_{\mathbf{k}}}{\int \varepsilon_{\mathbf{k}}^{r}f_{\mathrm{FD}}[1-f_{\mathrm{FD}}]d\varepsilon_{\mathbf{k}}}, \tag{32}$$

with an index *r* dependent on the properties of a specific material. Regarding 3D materials, we have $D_{\mathrm{OS}}\sim\varepsilon_{\mathbf{k}}^{1/2}$ and $v_{\mathrm{e}}\sim\varepsilon_{\mathbf{k}}^{1/2}$. As for the relaxation time, it depends on the scattering types, summarized in Tal. 1.

According to Eq. (32), we can estimate the lower and upper bounds of the Seebeck coefficient band term: the reversible model corresponds to the lower bound with $r$ equal to 1/2, while the BTE model with the weakly-screened-ionized impurity scattering dominating (highly-doped around room temperature) gives the upper bound, that is, $r = 3$.

TABLE 1 Values of the exponent in the energy dependence for different scattering mechanisms in doped semiconductors [23].

| **Scattering type** | **Energy dependence** |
|---|---|
| Acoustic phonon | $\sim\varepsilon_{\mathbf{k}}^{-1/2}$ |
| Ionized impurity (strongly screened) | $\sim\varepsilon_{\mathbf{k}}^{-1/2}$ |
| Neutral impurity | $\sim\varepsilon_{\mathbf{k}}^{0}$ |
| Piezoelectric | $\sim\varepsilon_{\mathbf{k}}^{1/2}$ |
| Ionized impurity (weakly screened) | $\sim\varepsilon_{\mathbf{k}}^{3/2}$ |

# III. Model Implementation and Comparison in n-Doped Silicon (Si)

Our investigation above is conducted in a theoretic view, while the three models are here applied to highly-n-doped Si (the doped concentration $n_\mathrm{d} > 1\times10^{19}$ cm$^{-3}$) [12][13][20][32] as a case study, where the band term is recognized to be predominant. The energy band, DOS, and electron velocity, Fermi level of Si are calculated following our previous paper [6]. Specifically, the relaxation time is assumed to be dominated by the weakly-screened-ionized impurity scattering that follows $\tau_\mathrm{e}\sim\varepsilon_{\mathbf{k}}^{3/2}$, which estimates the upper bound of Seebeck coefficient.

Figure 3 compares the absolute Seebeck voltage $|T\alpha_\mathrm{S}|$ values of n-doped Si samples measured by the experiments of Geballe et la. [12] ($n_\mathrm{d} = 2.7\times10^{19}$ cm$^{-3}$, bulk) and Sadhu et al. [20] ($n_\mathrm{d} = 6.0\times10^{19}$cm$^{-3}$, bulk; $n_\mathrm{d} = 7.0\times10^{19}$cm$^{-3}$, nanowire) with those predicted by the (a) reversible model ($r = 1/2$), (b) ballistic model ($r = 1$), and (c)

BTE model ($r$ = 3). According to Fig. 3(a), the reversible model's predictions generally serve as the lower bounds of Seebeck voltage for all the three doping concentrations; the mean relative deviation for $2.7\times10^{19}$ $cm^{-3}$ is 13%, that for $6.0\times10^{19}$ $cm^{-3}$ is 18%, and that for $7.0\times10^{19}cm^{-3}$ is 5%. The reversible model predictions are fairly good, considering its very concise form and low computational cost, which indicates this model can work well in the near-equilibrium case as clarified above. Moreover, the experimental data for $6.0\times10^{19}$ $cm^{-3}$ is significantly higher than that for $7.0\times10^{19}cm^{-3}$; for instance, at around 350 K, the Seebeck voltage for $6.0\times10^{19}$ $cm^{-3}$ is about 20% higher than that for $7.0\times10^{19}cm^{-3}$, which is unconventional, since they hold much close doping levels (a minor difference of Seebeck voltage, less than 3%, is predicted by all the three models) and are both dominated by the band term that should not be influenced by the quench of phonon-drag term due to the shrunk size in the $7.0\times10^{19}cm^{-3}$ case (nanowire) [20].

Figure 3 (b) presents the predictions by the ballistic model, which may be either higher or lower when comparing to the identical experimental data. For this model, the mean relative deviation for $2.7\times10^{19}$ $cm^{-3}$ is 3%, that for $6.0\times10^{19}$ $cm^{-3}$ is 6%, and that for $7.0\times10^{19}cm^{-3}$ is 16%: the prediction performance for the two formers is improved, while it overestimates the data for the last case.

Regarding the BTE model with weakly-screened-ionized impurity scattering, as shown in Fig. 3(c), it gives the significant overestimations in all the three cases: the mean relative deviation for $2.7\times10^{19}$ $cm^{-3}$ is 60%, that for $6.0\times10^{19}$ $cm^{-3}$ is 61%, and that for $7.0\times10^{19}cm^{-3}$ reaches 90%. This suggests the importance of setting reasonable relaxation time formulation when using the BTE model for calculating the Seebeck coefficient, which causes the major barriers and computational cost in the practical applications.

Furthermore, a parametric study of the three models is conducted, and Figure 4 gives the model predictions about the absolute Seebeck voltage of n-doped Si varying with doping concentration at different temperatures. All the models' results can increase with the reduced doping concentration or the enhanced temperature, and the reversible model always provides the lower bound predictions.

## IV. Concluding Remarks

The present work revisits theoretical modeling of Seebeck coefficient, with presenting some thermodynamic flaws in the existing methodologies: the confusion of electrochemical potential and electrical potential in the definition, arbitrarily neglecting the dependence on conduction band bottom and Fermi level when evaluating the spatial gradient of distribution function in the BTE, and improper heat flux presentation.

Hence, we propose an alternative methodology to derive the closed-form model of Seebeck coefficient in the near-equilibrium state, through using the drift-diffusion equation with the Soret effect based on the OC condition, which avoids the flaws above. In the near-equilibrium case, the reversible model [5] is also valid for the band term and gives the lower bound of Seebeck coefficient in theory, though it eliminates the velocity and relaxation terms; by contrast, the phonon-drag term is of the identical form to the BTE-based one.

Moreover, in the case study of highly-n-doped Si, the reversible, ballistic and BTE models are investigated and compared, along with the experimental data. It is found that the reversible model can provide the fairly-good estimations for the experiments. Considering its theoretical-soundness, simple form and low computation cost, it can serve as a fast and concise alternative for evaluating the TE properties of materials, which should be particularly useful for high-throughput calculations.

## Acknowledgements

None.

## Figure captions

Figure 1 Band diagrams of electron system (n-doped semiconductors) at (a) low temperature $T_c$ and (b) high temperature $T_h$, respectively: $E_C$: conduction band; $E_V$: valence band; $E_F$: Fermi level; $\mu$: chemical potential.

Figure 2 Two heat sinks are connected by a n-doped semiconductor in the condition of open circuit with the local equilibrium assumption (LEA); (b) Two heat sinks are connected by a n-doped semiconductor in the condition of open circuit in ballistic case.

Figure 3 Comparison between experimental data [12][20] of absolute Seebeck voltage $|T\alpha_{\mathrm{S}}|$ of n-doped Silicon samples and model predictions: (a) Reversible model $|T\alpha_{\mathrm{S(R)}}|$; (b) Ballistic model $|T\alpha_{\mathrm{S(Ba)}}|$; (c) BTE model $|T\alpha_{\mathrm{S(BTE)}}|$.

Figure 4 Model predictions for absolute Seebeck voltage $|T\alpha_{\mathrm{S}}|$ of n-doped Silicon samples varying with doping concentration ($n_d$) at different temperatures.

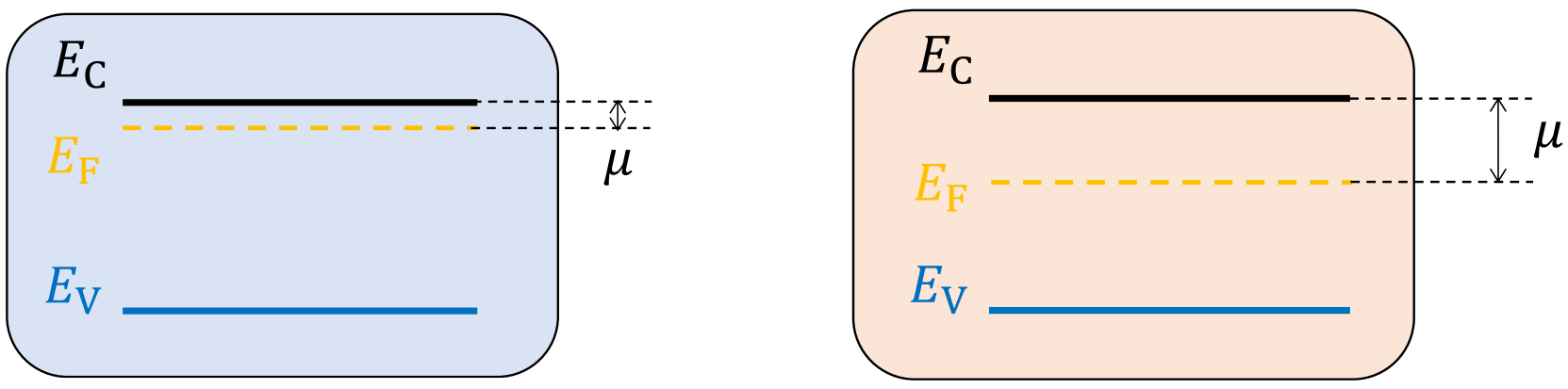

Figure 1 Band diagrams of electron system (n-doped semiconductors) at (a) low temperature $T_c$ and (b) high temperature $T_h$, respectively: $E_C$: conduction band; $E_V$: valence band; $E_F$: Fermi level; $\mu$: chemical potential.

(a) Two heat sinks are connected by a n-doped semiconductor in the condition of open circuit with the local equilibrium assumption

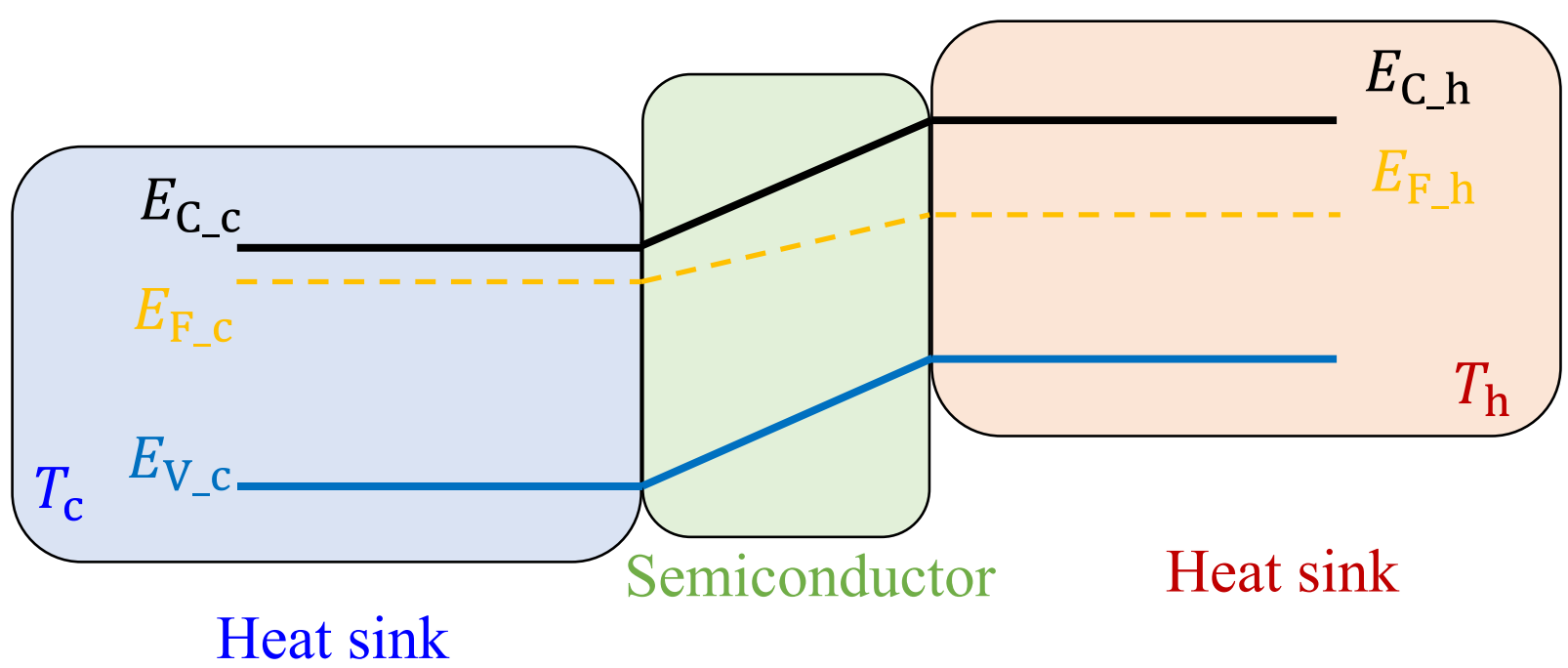

(b) Two heat sinks are connected by a n-doped semiconductor in the condition of open circuit in ballistic region

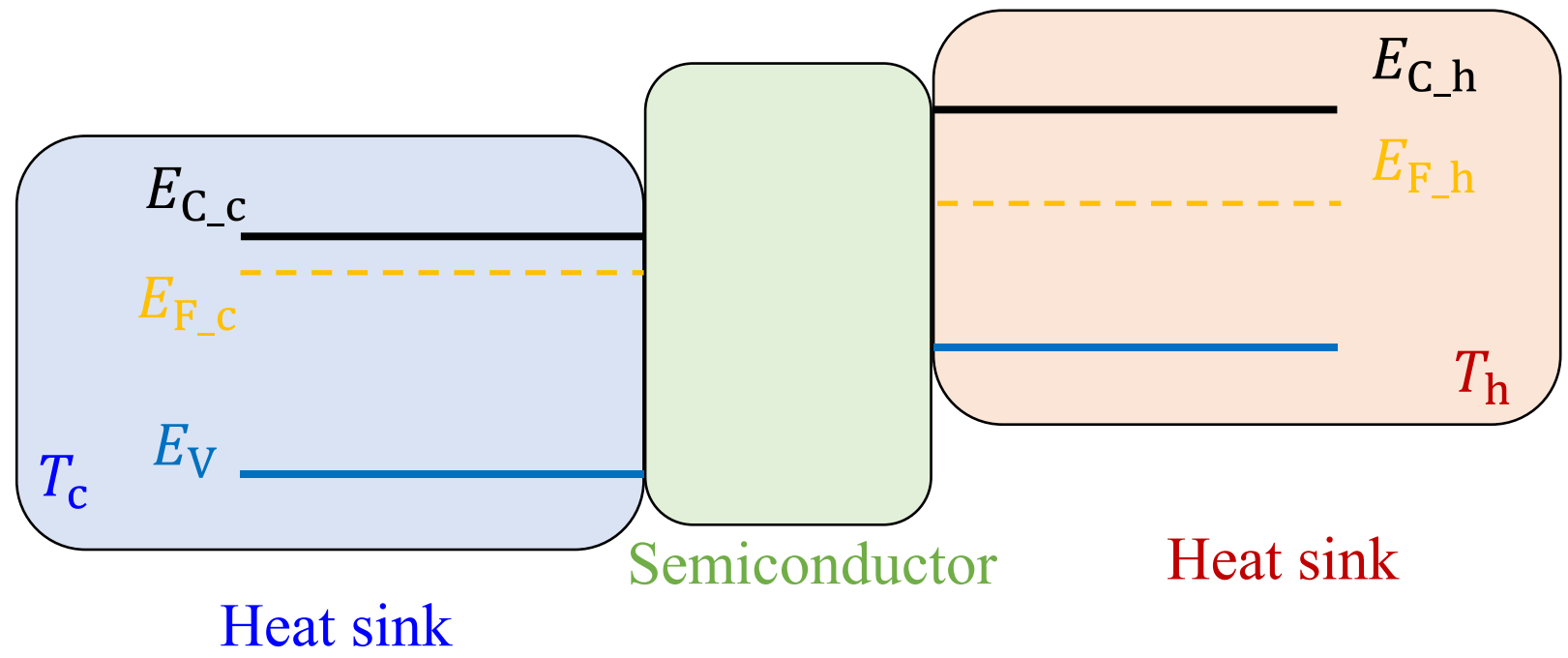

Figure 2 (a) Two heat sinks are connected by a n-doped semiconductor in the condition of open circuit with the local equilibrium assumption (LEA); (b) Two heat sinks are connected by a n-doped semiconductor in the condition of open circuit in ballistic case.

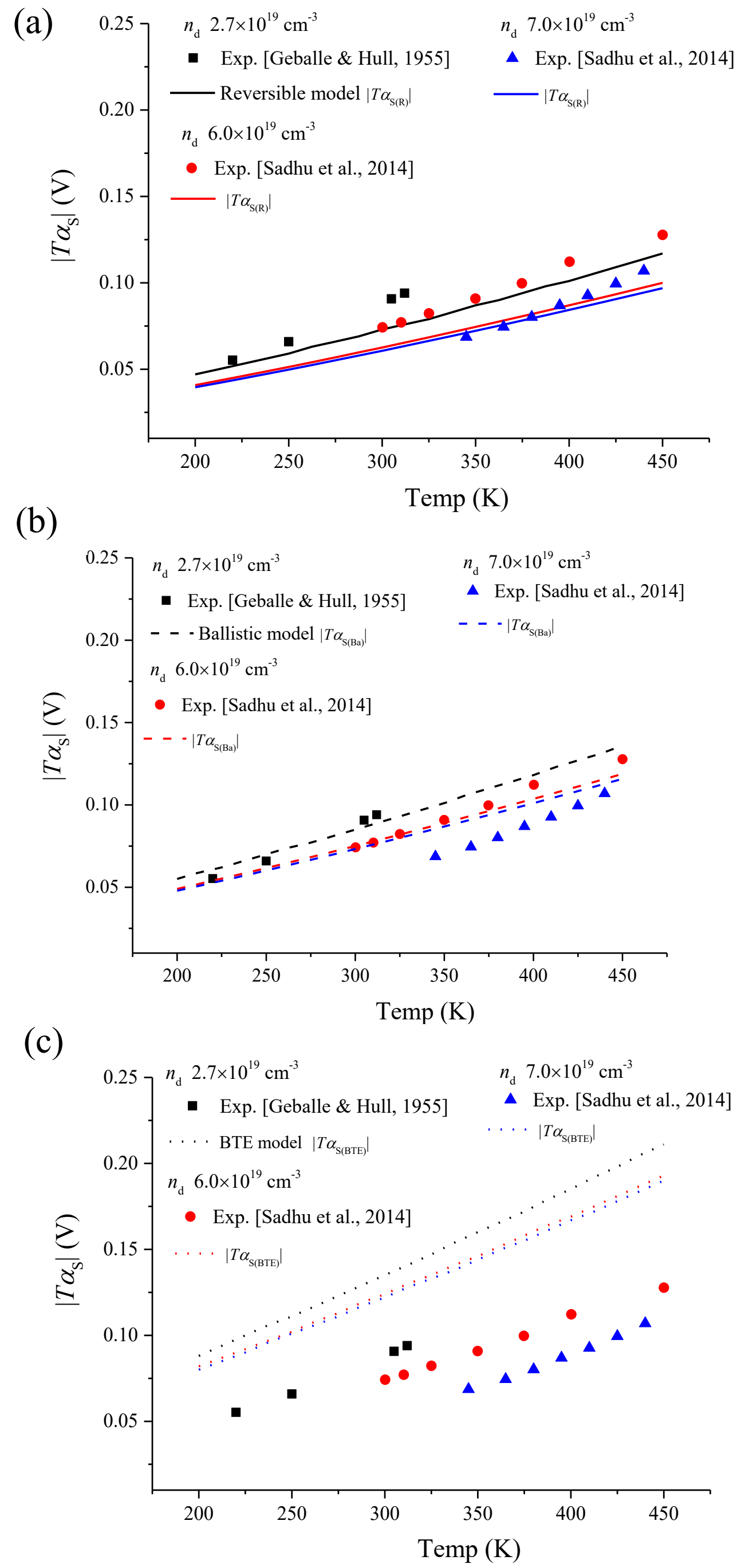

Figure 3 Comparison between experimental data [12][20] of absolute Seebeck voltage $|T\alpha_{\mathrm{S}}|$ of n-doped Silicon samples and model predictions: (a) Reversible model $|T\alpha_{\mathrm{S(R)}}|$; (b) Ballistic model $|T\alpha_{\mathrm{S(Ba)}}|$; (c) BTE model $|T\alpha_{\mathrm{S(BTE)}}|$.

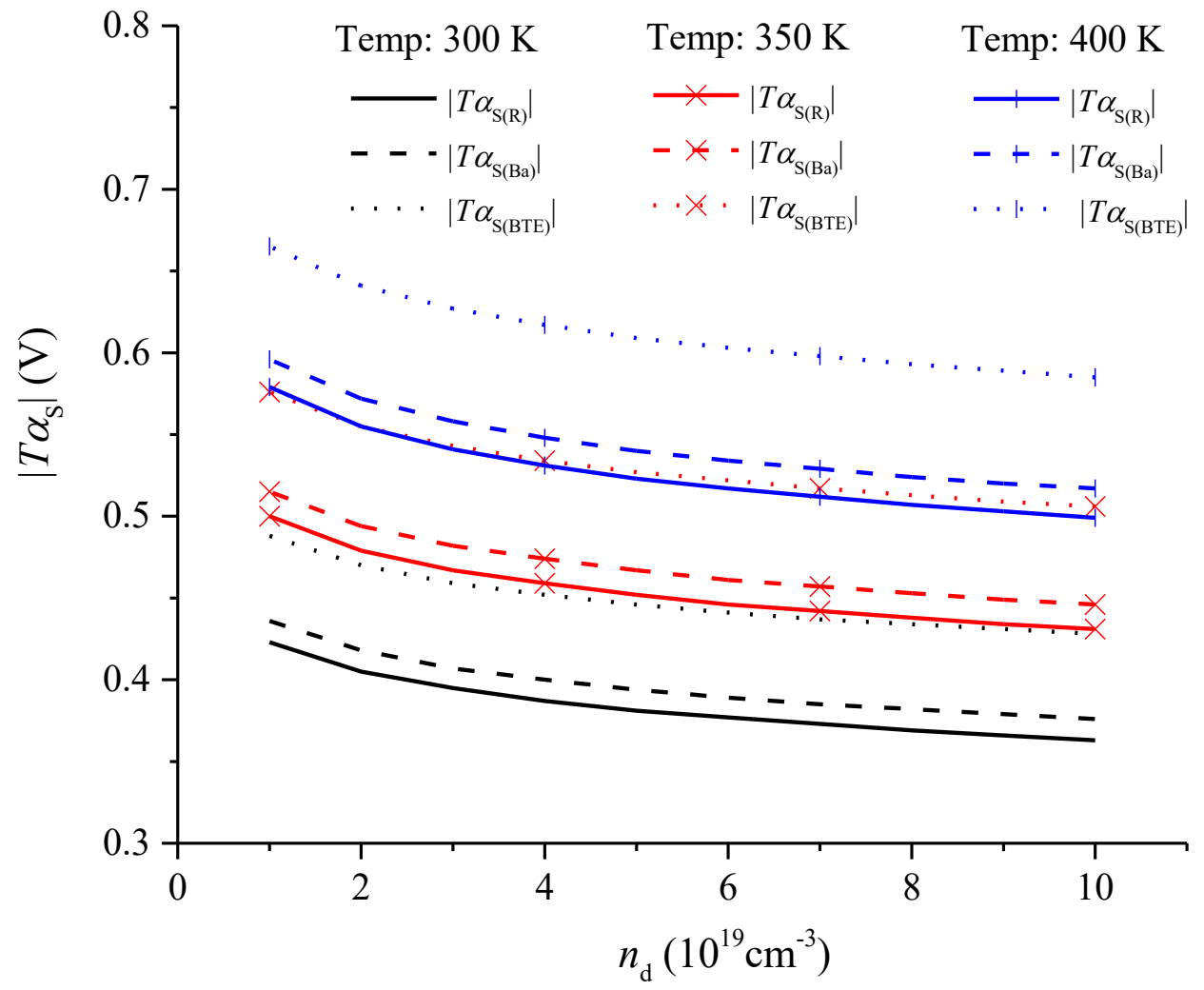

Figure 4 Model predictions for absolute Seebeck voltage $|T\alpha_S|$ of n-doped Silicon samples varying with doping concentration ($n_d$) at different temperatures.